# The Twilight of Determinism: At Least in Biophysical Novelties

Amihud Gilead
Department of Philosophy, Eshkol Tower, University of Haifa, Haifa 3498838
Email: agilead@research.haifa.ac.il

*Abstract*. In the 1990s, Richard Lewontin referred to what appeared to be the twilight of determinism in biology. He pointed out that DNA determines only a little part of life phenomena, which are very complex. In fact, organisms determine the environment and vice versa in a nonlinear way. Very recently, biophysicists, Shimon Marom and Erez Braun, have demonstrated that controlled biophysical systems have shown a relative autonomy and flexibility in response which could not be predicted. Within the boundaries of some restraints, most of them genetic, this freedom from determinism is well maintained. Marom and Braun have challenged not only biophysical determinism but also reverse-engineering, naive reductionism, mechanism, and systems biology.

Metaphysically possibly, anything actual is contingent.[1] Of course, such a possibility is entirely excluded in Spinoza's philosophy as well as in many philosophical views at present. Kant believed that anything phenomenal, namely, anything that is experimental or empirical, is inescapably subject to space, time, and categories (such as causality) which entail the undeniable validity of determinism. Both Kant and Laplace assumed that modern science, such as Newton's physics, must rely upon such a deterministic view. According to Kant, free will is possible not in the phenomenal world, the empirical world in which we exist and which is fully deterministic, but only in the noumenal— the morally rational realm—in which we are free not to follow our emotions and instincts but only the moral, absolute law whose legislator is solely our reason. In contrast, according to Spinoza—who completely excluded the possibility of free will—contingency is only a matter of our ignorance; whenever we know the causes that are relevant to any change, we know for sure, quite adequately, that it does not happen contingently.

---

[1] This is an assumption of a full-blown metaphysics—panenmentalism—whose author is Amihud Gilead. See Gilead 1999, 2003, 2009 and 2011, including literary, psychological, and natural applications and examples —especially in chemistry —in Gilead 2009, 2010, 2013, 2014a–c, and 2015a–d.

Many believed that each of the natural sciences demonstrates that empirical reality, perhaps the only reality in which we can live, is deterministic, as nothing happens contingently, as everything is subject to strict causality (quantum mechanics makes no exception). This holds true not only for all physical, chemical, and biological ordinary phenomena but also for chaotic ones. Chaotic phenomena—those phenomena whose equilibrium is extremely sensitive to the tiniest difference in the initial conditions and hence is unstable—are subject to mathematical equations. Thus, even though we cannot make weather forecasts for a long period but only for the next three or four days, any weather prediction is subject to simple mathematical equations. Hence, such chaotic phenomena are deterministic.

One can agree that when astronomical phenomena are concerned, determinism appears to have quite a solid ground (some will doubt even this). However, what about biology? Is life, its very existence, subject to determinism, even subject to it at all? Think of Darwinism—there are evolutionary laws but all evolutionary changes cannot escape the contingency of circumstances that are not biological, genetically or otherwise. Nevertheless, also in this case, one can take a Spinozistic stance and argue that all this "contingency" is nothing but our ignorance about these circumstances in which, too, nothing happens contingently.

Though panenmentalist metaphysics leaves necessity only to the realm of individual pure possibility, actual-physical reality is contingent only, even though any actuality is an incomplete actualization of an individual pure possibility (see Gilead's panenmentalist publications mentioned above). For a clarifying analogy, think of purely mathematical entities. These entities or objects yield to a strict logico-mathematical necessity, which is *a priori*, and, yet, the case appears to be that these entities or objects and their necessary relations are actualized contingently only as physical actualities. Nothing could enforce them as necessary on actual reality. May we demonstrate this to be the case in biological phenomena, too? Is such the case in these phenomena not because of our ignorance but because it is so in fact, quite independent of our knowledge?

In the 1990s, discussing biological phenomena, Richard Lewontin (1995 and 2000) argued that their major parts did not yield to determinism, especially those associated with genetics. Life phenomena are most complex, and genetics plays a role in them but not a crucial or major one. The interrelations between any organism and its environment are even more critical. Lewontin writes:

> Even the organism does not compute itself from its DNA. A living organism at any moment in its life is the unique consequence of a development history that results from the interaction of and determination by internal and external forces. . . . Organisms do not find the world in which they develop. They make it. . . . Nor is "internal" identical with "genetic". . . . The variation between sides is a consequence of random cellular movements and chance molecular events within cells during development, so-called "developmental noise". It is this same developmental noise that accounts for the fact that identical twins have different fingerprints and that the fingerprints on our left and right hands are different. . . . The scientists writing about the Genome Project explicitly reject an absolute genetic determinism, but they seem to be writing more to acknowledge theoretical possibilities than out of conviction. . . . we cannot really believe that the sequence of the human genome is the grail that will reveal to us what it is to be human. (Lewontin 1995, pp. 61–62; cf. 2000, p. 120)

Following the knowable and wise advice of Lewontin, biology should not dispense of contingency, unpredictability, and indeterminism—all which are inevitable for knowing and understanding biology as it really is.

What really matters, according to such a view, is the flexibility to operate without a "program"; it is the capability of accommodating unforeseen challenges (to use Braun's phrase for it).

Indeed, two biophysicists, Shimon Marom and Erez Braun from the Technion at Haifa, Israel, have clearly demonstrated that a population of microorganisms (such as yeast) and a population of neurons are not subject to determinism, let alone to a strict one; instead, such populations demonstrate degrees of spontaneity and freedom which are unpredictable in two senses—reversible and progressive (Braun and Marom 2015; Braun 2015). Such spontaneity and freedom are limited by some initial conditions but not by any sort of determinism, let alone a strict one. Implying much philosophical significance, this enlightening novelty does not ignore genetics, but it certainly shows that there are *pure possibilities* that are open to such systems of organisms and which are actualized in a contingent, or at least, indeterministic ways.

In a new, exciting and enlightening book, Marom shows that this kind of freedom and indeterminism is valid for the states and activity of our brain (Marom 2015). Marom and Braun's scientific findings challenge the philosophies of systems biology (especially in rendering organisms, first of all our brain, as subject to mathematical models regardless of the environment in which our brain exists and operates), reductionist reverse-engineering, determinism, mechanism, and functionalism.[2] It is a pity that about 27 years since Hilary Putnam's philosophical candid and admired confession, that even though he was an ardent functionalist in the 1970s, since 1988 he has relinquished functionalism entirely (compare Putnam 1977 to Putnam 1988). Marom points to the major deficiency of functionalism in a most beautiful example (Marom 2015, p. 29–30): Garry Kasparov's complaint that chess players start to imitate the Deep Blue computer instead of relying upon their insights and ingenuity (which reminds me of the famous scene in Chaplin's *Modern Times*).

---

[2] For a reductionist approach to psychology and natural science, one of which is biology, reducing the macro-reality to the physical micro-reality, see Hemmo and Shenker 2015, explicitly mentioning the "determinism at the fundamental mechanical level" (ibid., p. 6). Such a deterministic view is quite prevalent nowadays. Note that Hemmo and Shenker do not refer to neurons or cells but to quantum mechanics, thus the reductionism they have in mind is quite complicated and much more fundamental than the alternative physicalist views.

I am convinced that Marom and Braun's challenge is well established and should not be dismissed or ignored.

Though some philosophers of science have challenged the indeterministic interpretation of Braun and Marom's findings (see Carrier and Hon 2015), I believe that these philosophers have demonstrated a dogmatic, even a blind attitude, which refuses to learn anything new from established scientists, just because their novel discoveries are not compatible with the dogmatic assumptions of such philosophers. It is not in my intention to explicate my detailed counter-arguments to their criticism but, instead, to do my job as a philosopher and metaphysician, namely, to analyze open-mindedly the philosophical significance of these biophysical findings instead of criticizing them simply on a dogmatic basis. It is the job of *natural scientists* to put such findings to test, not of *philosophers.* It is a pity that those philosophers who criticizing Marom and Braun on some philosophical grounds, simply forgot or ignored the great Humean lesson that even though Newtonian physics was incompatible with Hume's empiricism regarding his criticism of necessary connections such as causality, Hume never doubted the great achievements of Newtonian physics. The adequate scientific test is most of all empirical. Philosophers are not allowed to criticize natural science merely on philosophical grounds. Philosophers should not consider themselves as methodologists of natural sciences nor critics of it. They should be students of science to the extent that empirical conclusions or theoretical ones are concerns. In contrast, the case appears to be that the philosophical group at Konstanz,[3] who criticized Marom and Braun, simply forgot what philosophers should do in studding a new scientific approach. Out of a philosophical self-assurance or dogmatism, which has not much to do with candid and open-minded philosophy, they appear to react like dogmatic clerics, blind to facts and fresh innovative scientific ideas.

In an opposite way, I find some major similarities between panenmentalism and Marom and Braun's biological approach. Marom writes: "Analyses in terms of relational dynamics with the environment . . . are doomed to fail in the short term because . . . they are too general and not obviously applicable, they are romantic and abstract, and their proponents are willing to accept the *impossible* as valid no less than the *possible*" (Marom 2015, p. 174). It is quite easy to translate this into panenmentalist terms thus: instead of "actually impossible" panenmentalist mentions "purely possible" that is quite valid. Furthermore, Braun's "contribution . . . to active protection against biological determinism" (ibid.)[4] is certainly compatible with the panenmentalist conclusion that empirical reality in general and the biological one in particular are not deterministic, as anything actual is contingent. Panenmentalism leaves necessity only to the realm of individual pure possibilities and their relationality (namely, the ways in which they relate to each other). Equally, applied mathematics deals with contingency, for there is no reason that one of the mathematical pure possibilities should be actualized instead of another one, whereas in each of the fields of pure mathematics necessity inescapably prevails.

---

[3] Namely, Carrier and Hon 2015, Bechtel 2015, Green 2015, and Krohs (2015).

As for Braun's insightful conclusions from his experiments with yeast, conclusions that *are valid also for the formation of cancer*, he writes: "In the language based on the lessons from our yeast experiments . . . in the context of differentiation, the cell realizes its basic *potential* once triggered into an *exploratory* mode; within *the broad spectrum of possible realizations*, parallel to differentiated states in normal developmental trajectories *there are other trajectories that lead to*

---

[4] Cf. the following: "The genotype, notwithstanding its important role, participates in the process but *does not fully determine it*. The genome in this view provides a set of constraints on the spectrum of regulatory modes, analogous to boundary conditions in physical dynamical systems" (Braun 2015, p. 44; italics are mine).

*cancer*" (Braun 2015, p. 41; italics are mine).[5] Simply replace the word "potential" by "individual pure possibility", and you traverse the line distinguishing natural science and panenmentalism as a metaphysics. It is simply a matter of translation—translating the natural scientific findings and insights into metaphysical insights, concepts, and explanations.

Note that Amihud Gilead discovered and was aware of all the natural scientific applications or correlates of panenmentalism, after, never before, he had originated and elaborated on this metaphysics. The first panenmentalist application was to consider the chemical elements, including the eka-elements, in the periodic table as the upshot of the attempt to *exhaust all chemical pure possibilities*. Gilead realized this application only after the publication of his first panenmentalist work (Gilead 1999) while writing the second panenmentalist book (Gilead 2003, see pp. 65–70). Not before 2012, did Gilead discover the second application, that of the quasicrystals, to which an exclusion of vital crystalline pure possibilities had preceded by Pauling and his many followers. Gilead discovered this major and most interesting application only after the publication of his fourth panenmentalist book (in 2011). The same holds for some other revolutionary discoveries in chemistry, physics, and biology to which he devoted some of his papers. The same holds true for the findings of Marom and Braun, to which he has become familiar only very recently.

---

[5] Cf. "every cell has the potential to adapt via multiple heterogeneous processes" (op.cit. p. 40) and "The point in cancer is that in the context of the adult organism, initiating stemness and novel trajectories of differentiation may lead to disastrous consequences in an inappropriate context while in the evolutionary context it serves as a potential for innovation and facilitates evolvability" (op. cit., p. 41). In each of these cases, saving possibilities—a panenmentalist credo and the title of the first panenmentalist book (Gilead 1999)—plays in fact a major role.

How is it possible that a metaphysician, a philosopher who relies upon *a priori* considerations and entertains pure possibilities and their relationality, had discovered the relevant pure possibilities of these scientific, empirical discoveries? The reason is that philosophers, poets, authors, artists, mathematicians, as well as natural scientists *all have some access to the realm of pure possibilities*. They do not necessarily need to create models to make their actual discoveries, they need their reasoning and especially imagination to have access to the realm of pure possibilities, which, unlike empirical facts, are not accessible by relying upon experience and observations. Pure possibilities, by definition, are not empirically observable or discoverable. In the same manner, Platonists in the field of the philosophy of mathematics refer to the imagination and reasoning of mathematicians as legitimate means to have access to the realm of the entities or objects of pure mathematics. As much as philosophers may open our eyes to pure possibilities that we have never encountered as actualities, dogmatic philosophers may attempt to close such possibilities for us. In fact, such was the manner that the above-mentioned Konstanz group of philosophers of science had treated the innovations of Marom and Braun.

The following is a crucially important passage:

> . . . protein fluctuations do not reflect any specific molecular or cellular mechanism, and suggest that some buffering process masks these details and induces universality . . . In the neural system, the same feature is manifested in long-term single neuron and neuronal population excitability dynamics, which are unstable and dominated by critical fluctuations, intermittency, scale-invariant rate statistics, and long memory processes. . . . Physics teaches us that such broad, universal distributions that emerge in different systems and scales, require either fine tuning of control parameters 'engineered' to lock the system in specific (but rare) points in phase space (this is then a critical point), or a capacity of the system to tune itself to hover around such a point, irrespective of control parameters (a process coined self-organized criticality). (Braun and Marom 2015, p. 3)

Referring to a capacity of the system *to tune itself* to hover around such a critical point is a new, overwhelming voice in biophysics. Mechanisticness, computabilism, functionalism, and most notably determinism, ignoring the inter-relationality of complex systems and environment, should not be considered from now on as suitable or adequate in order to learn and understand such biophysical systems. Indeed, thanks to Marom and Braun's novel view, "the path towards complexity is wide open" (ibid.). Hence,

> it turns out that the microorganism system can adapt to unforeseen challenges within a few generations, practically instantaneous in evolution terms; moreover, the rapidly emerging adapted state is stably inherited . . . This is analogous to the facts of fast learning and memory in humans and other organisms, attributed to whole brain mechanisms. Such rapid adaptations, learning and memory that seemingly break efficacy limits imposed by the dimensionality of the problems, are also observed in our large-scale neuronal networks. . . . The above examples of microscopic-macroscopic degeneracy of observables led us, in Konstanz, to reflect on the *possible indeterminacy* entailed by the combination of many-to-one and one-to-many relations among

> levels of organization: Regardless of the level one chooses to analyze, the extent to which observables from that analyzed level determine the phenomenology at other levels, seems limited. We suspect that one possible origin of the above microscopic-macroscopic degeneracy is related to *our (experimentalists) habitual isolation of parts from the whole in standard experimental praxis*. Biological systems under natural conditions are embedded in environments, which are in themselves dynamical entities that mould the—and are coupled to—many levels of system organization, from the single cell to the whole organism and population of organisms. *Disconnecting* the system's dynamics from the dynamics or statistics of the environment, might lead to *erroneous classification of system's phenomena*. A demonstration of the latter point involves the interpretation of neuronal response variability under different environmental statistics. . . . While the exposure of universals is a most significant aspect of biological research (e.g. DNA, cell structure, energy production and consumption, etc.), *much of biology is about specificity*, telling the origins of differences between species, phenomena, and capacities. . . . it implies that *the search for the relevant system variables should become the most important and urgent for the advancement of biology. Without identification of the relevant variables, the practice of experimental biology becomes a fishing expedition, dictated by fashion and technological barriers*. As demonstrated below, reverse engineering of biological systems constitutes an example for going amiss upon lack of well defined relevant system variables. (op. cit., pp. 3–4; the italics are mine)

In this way, Marom and Braun have opened new, revolutionary possibilities, at least for biophysics. A panenmentalist approach can serve them, as well as scientists like them, as a philosophical platform or frame for their fruitful research whose future is certainly still ahead of the present era. Especially because of biases, prejudices, and dogmatism, some of which, unfortunately, have had some philosophical support. Panenmentalism aims at saving pure possibilities, which are vital for our knowledge, conduct of life, the progress of natural sciences, or the very existence of arts. Pure possibilities are essential for our capability to use our imagination and creativity, which has to do with one of our special feature—neoteny—which mainly consists of our playing or entertaining with pure possibilities (Gilead 2015d). Following Donald Wood Winnicott's view of playing, we are well aware how vital is the role that it plays in scientific creativity and discoveries, no less than in philosophy and the arts. Panenmentalism may serve as a conceptual frame or philosophical basis for Marom and Braun's scientific innovations not only in saving new scientific pure possibilities that the mainstream of biophysics has neglected so far or even has closed or excluded, but also in regarding actual-physical reality as contingent, hence in rejecting or, at least, drastically mitigating determinism to the extent that biology is concerned. Panenmentalism also gives strong support to their innovations by rejecting reductionism that is not naive (to use their term). Panenmentalism clearly explicates the difference of categories between pure possibility and its physical actuality (physical in the broad sense of the term, i.e., strictly physical, material, chemical, biochemical, or biological). Pure possibilities are absolutely exempt from spatiotemporal and causal conditions as well as from any physical properties, whereas actualities inescapably yield to spatiotemporal and causal conditions as well as to physical properties. For instance, actualities (namely, actual entities) have mass or weight, they have a shape or figure, have color, size, dimensions, temperature, chemical properties, and the like, whereas pure possibilities are exempt from all these. Because actualities are thus caterorially different from their essential pure possibilities (which determine the identity of these actualities), they are entirely irreducible to each other. Furthermore, according to panenmentalism, biological entities, which actualize biological pure possibilities, are irreducible to physical entities, which actualize physical pure possibilities (as their identities) or to chemical or biochemical entities, which actualize chemical or biochemical pure possibilities (as their identities). Thus, organisms—biological entities—which actualize biological pure possibilities, are irreducible to physical, chemical, or biochemical pure possibilities. Each actuality actualized its own pure identity-possibility, not that of another actuality. All these pure possibilities are discoverable by means of the theory in each of the relevant scientific fields. If, because of a dogmatic attitude, some of these essential possibilities are excluded, the scientific progress is blocked and scientists could not identify new, so-far unfamiliar, phenomena and could not be able to identify them. In all of these senses, panenmentalism sets the conceptual frame or the philosophical-metaphysical basis for new scientific ideas such as those of Marom and Braun.

The time is ripe to relinquish in some natural sciences, such as biology, the deterministic nature that Kant and Laplace ascribed to all natural phenomena. If such is the case in biophysics, all the more so is the case in biology in general. At least in biology in general and brain studies in particular, despite the prevailing of main stream reverse engineering, functionalism, computabilism, and mechanism,[6] the twilight of determinism begins to raise its much promising head.

A good philosophy of science should be an analytical and detailed *reflection* on scientific novelties, discoveries, and progress; it should never take upon itself the task of a mentor, compass, or regulator. Such is the manner that suits religious tendencies or, worse, politically totalitarian, not of a genuine, good philosophy. It is about time to liberate our ideas of natural science from such a dogmatism that such group of philosophers of science in Konstanz have, most unfortunately, demonstrated. Indeed, "the multitude of possible mappings exposes the inherent difficulty of reverse engineering – that is, its indeterminacy" (op. cit., p.4), as complexity is indispensable for biology; complexity is the special nature of biological systems.

Molecular biology, with the grand ambitions of James Watson and Francis Crick, has followed the wrong approach that all the biological phenomena lay on the basis of the DNA structure and its ramifications. This has turned out to be a fantasy, and a very misleading one. Braun and Marom offer biologists a paradigm-change, a revolution in theoretical biology. It is very sad to realize that of all scholars some philosophers are entirely blind to this promising and welcome novel, even revolutionary, shift.

---

[6] Indeed, "unlike technology, the business of biology as a basic science is not to uncover a plausible mechanism but rather to discover the actual design principles underlying the natural phenomenon; this is where the naive version of reverse engineering in particular, and naive reductionism in general, epistemically fails" (op. cit., p. 4). This is a firm *philosophical* reflection and not only a wise natural scientific guideline.

I would like to end this paper by citing the most important following promising hope:

> Notwithstanding the impressive advance in molecular biology, the last 50 years have taught us that progress in understanding biology, which is not synonymous with progress in medical applications, is severely impeded by the reductionist approach, focusing solely on cataloguing an ever increasing list of molecular processes, without a complementary effort in unraveling the system-level organization principles. What seems to be missing is indeed a unifying concept of organization. . . . The current multi-disciplinary effort is still on, so it might be too early to judge. However, until now it has not yet developed an original view, raising its own voice regarding the origin, evolution and development of biological systems as *natural phenomena,* independent of the tyranny of the molecular approach. (Braun 2015, p. 45)

As a philosopher and metaphysician, I whole-heartedly welcome this hope. I look forward to witnessing "a shift in conceptual thinking from molecular causations to a problem of *organization*" (Braun 2015, p.38).[7] Such conceptual thinking is of great interest to open-minded philosophers and scientists as well. It is about time for open-minded philosophers not to follow a reductionistic approach and the old-fashioned tyranny of the molecular approach in biology. It is about time to reconsider to what extent, if at all, determinism is valid at least for the biosciences.

---

[7] Cf. the "paradigm shift" in the study of cancer: "putting the environment context and history of the cells at central stage while moving genes backstage" (Braun 2015, p. 39); "a shift from the somatic mutation picture to *developmental-organizational* perspective" (op.cit., p. 40). It is very interesting and enlightening to compare this conception cancer with that of Aktipis et al. 2015.

References

Aktipis CA, Boddy AM, Jansen G, Hibner U, Hochberg ME, Maley CC, Wilkinson GS. (2015) "Cancer across the tree of life: Cooperation and cheating in multicellularity", *Philosophical Transaction of the Royal Society B* 370: 20140219. http://dx.doi.org/10.1098/rstb.2014.0219

Bechtel, William (2015) "Can mechanistic explanation be reconciled with scale-free constitution and dynamics?" *Studies in History and Philosophy of Biological and Biomedical Sciences* xxx, pp. 1–11. http://dx.doi.org/10.1016/j.shopsc.2015.03.006.

Braun, Erez and Marom, Shimon (2015) "Universality, complexity and the praxis of biology: Two case studies", *Studies in History and Philosophy of Biological and Biomedical Sciences* xxx, pp. 1–5. http://dx.doi.org/10.1016/j.shopsc.2015.03.007.

Braun, Erez (2015) "The unforeseen challenge: From genotype-to-phenotype in cell populations", *Reports on Progress in Physics* 78:3, pp. 1–51.

Carrier, Marin and Hon, Giora (2015) "Introduction: Philosophers meet biologists", *Studies in History and Philosophy of Biological and Biomedical Sciences* xxx, pp. 1–4. http://dx.doi.org/10.1016/j.shpsc.2015.04.001.

Gilead, Amihud (1999). *Saving Possibilities: An Essay in Philosophical Psychology* (Amsterdam & Atlanta: Rodopi – Value Inquiry Book Series, Vol. 80).

Gilead, Amihud (2003). *Singularity and Other Possibilities: Panenmentalist Novelties* (Amsterdam & New York: Rodopi – Value Inquiry Book Series, Vol. 139).

Gilead, Amihud (2005). "A possibilist metaphysical reconsideration of the identity of indiscernibles and free will," *Metaphysica* 6, pp. 25–51

Gilead, Amihud (2009). *Necessity and Truthful Fictions: Panenmentalist Observations* (Amsterdam & New York: Rodopi – Value Inquiry Book Series, vol. 202).

Gilead, Amihud (2010). "Actualist fallacies, from fax technology to lunar journeys," *Philosophy and Literature* 34:1, pp. 173–187.

Gilead, Amihud (2011). *The Privacy of the Psychical* (Amsterdam & New York: Rodopi – Value Inquiry Book Series, vol. 233).

Gilead, Amihud (2013). "Shechtman's three question marks: Impossibility, possibility, and quasicrystals." *Foundations of Chemistry* 15, pp. 209–224.

Gilead, Amihud (2014a). "Pure possibilities and some striking scientific discoveries", *Foundations of Chemistry* 16: 2, pp. 149–163.

Gilead, Amihud (2014b) "We are not replicable: A challenge to Parfit's view," *International Philosophical Quarterly* 54:4, pp. 453–460.

Gilead, Amihud (2014c) "Chain reactions, 'impossible' reactions, and panenmentalist possibilities", *Foundations of Chemistry* 16 (2014), pp. 201–214.

Gilead, Amihud (2015a) "Self-referentiality and two arguments refuting physicalism," *International Philosophical Quarterly* (forthcoming, December 2015).

Gilead, Amihud (2015b) "Can brain imaging breach our mental privacy?" *The Review of Philosophy and Psychology* 6:2, pp. 275–291.

Gilead, Amihud (2015c) "Cruelty, singular individuality, and Peter the Great," *Philosophia* 43:2, pp. 337–354.

Gilead, Amihud (2015d) "Neoteny and the playground of pure possibilities," *International Journal of Humanities and Social Sciences* 5:2, pp 30–39.

Green, Sara (2015) "Can biological complexity be reverse engineered?" *Studies in History and Philosophy of Biological and Biomedical Sciences* xxx, pp. 1–11. http://dx.doi.org/10.1016/j.shpsc.2015.03.008

Hemmo, Meir and Shenker, Orly (2015) "The emergence of macroscopic regularity", *Mind and Society*. DOI 10.1007/s11299-015-0176-x

Krohs, Ulrich (2015) "Can functionality in evolving networks be explained reductively?" *Studies in History and Philosophy of Biological and Biomedical Sciences* xxx, pp. 1–8. http://dx.doi.org/10.1016/ j.shpsc.2015.03.009

Lewontin, Richard (1995) *The Doctrine of DNA: Biology as Ideology* (Penguin).

Lewontin, Richard (2000) *The Triple Helix: Gene, Organism, and Environment* (Cambridge, MA: Harvard University Press).

Marom, Shimon (2015) *Science, Psychoanalysis, and the Brain: Space for Dialogue* (New York: Cambridge University Press).

Putnam, Hilary (1975) *Mind, Language, and Reality: Philosophical Papers*, vol. 2 (Cambridge, England: Cambridge University Press).

———. (1988) *Representation and Reality* (Cambridge, Mass.: MIT-Bradford).